
 \input phyzzx

\def\CMP#1{{\sl Comm. Math. Phys. {\bf #1}}}
\def\IMPA#1{{\sl Int. J. Mod. Phys. {\bf A#1}}}

\def\JPA#1{{\sl J.\ Phys.\ {\bf A#1}}}

\def\NPB#1{{\sl Nucl.\ Phys.\ {\bf B#1}}}

\def\PLB#1{{\sl Phys.\ Lett.\ {\bf #1B}}}

\def\TMP#1{{\sl Theor.\ Math.\ Phys.\ {\bf #1}}}


\def\nxl{\hfill\break}

\def\B{{\cal B}}

\def\G{{\cal G}}                            
\def\H{{\cal H}}                            
\def\T{{\cal T}}                            
\def\V{{\cal V}}                            

\def\a{\alpha}

\def\b{\beta}
\def\g{\gamma}

\def\l{\lambda}

\def\s{\sigma}
\def\om{\omega}
\def\t{\theta}

\def\Th{\Theta}

\def\o{\over}

\def\bold#1{\setbox0=\hbox{$#1$}
     \kern-.025em\copy0\kern-\wd0
     \kern.05em\copy0\kern-\wd0
     \kern-.025em\raise.0433em\box0 }
\def\lowmp{\lower.11em\hbox{${\scriptstyle\mp}$}}

\def\Im{{\rm Im\,}}

\def\sign{{\rm sign}}
\def\sub#1#2{{{#1^{\vphantom{0000}}}_{#2}}}
\def\frac#1#2{{\textstyle{
 #1 \over #2 }}}                            


\def\ZZ{{\rm Z \!\! Z}}                       
\def\CC{{\rm | \! \! C}}                    
\def\1{{\rm 1 \!\!\, l}}                        
\def\bra#1{\left\langle #1\right|}               
\def\ket#1{\left| #1\right\rangle}               
%

%
%


\hyphenation{Di-par-ti-men-to}
\hyphenation{na-me-ly}
\hyphenation{al-go-ri-thm}
\hyphenation{pre-ci-sion}
\hyphenation{cal-cu-la-ted}

%

%

\Pubnum={$\rm PAR\; LPTHE\; 93/07 \quad UPRF-93-370
         \qquad {\rm February \; 1993}$}
\date={}
\titlepage
\title{
YANG--BAXTER SYMMETRY IN INTEGRABLE MODELS: NEW LIGHT FROM THE BETHE
ANSATZ SOLUTION
\foot{Work supported by the CNR/CNRS exchange program}  }
\author{ C. Destri }
\address{ Dipartimento di Fisica, Universit\`a di Parma,
          and INFN, Gruppo Collegato di Parma
     \foot{E--mail: destri@parma.infn.it \nxl
           mail address: \nxl
           Dipartimento di Fisica \nxl
           Viale delle Scienze, 43100 Parma, ITALIA }}
\author{ H.J. de Vega }
\address{ Laboratoire de Physique Th\'eorique et Hautes Energies
     \foot{Laboratoire Associ\'e au CNRS UA 280}, Paris
     \foot{E--mail: devega@lpthe.jussieu.fr \nxl
           mail address: \nxl
           L.P.T.H.E., Tour 16, $1^{\rm er}$ \'etage, Universit\'e Paris
VI,\nxl
           4, Place Jussieu, 75252, Paris cedex 05, FRANCE }}

\vfil
\abstract
We show how any integrable 2D QFT enjoys the existence of infinitely many
non--abelian {\it conserved} charges satisfying a Yang--Baxter symmetry
algebra.
These charges are generated by quantum monodromy operators and provide a
representation of $q-$deformed affine Lie algebras.
We review and generalize the work of de Vega, Eichenherr and Maillet on the
bootstrap construction of the quantum monodromy operators to the sine--Gordon
(or massive Thirring) model, where such operators do not possess a classical
analogue. Within the light--cone approach to the mT model, we explicitly
compute the eigenvalues of the six--vertex alternating transfer matrix
$\tau(\l)$ on a generic physical state, through algebraic Bethe ansatz.
In the thermodynamic limit $\tau(\l)$ turns out to be a two--valued periodic
function. One determination generates the local abelian charges, including
energy and momentum, while the other yields the abelian subalgebra of the
(non--local) YB algebra. In particular, the bootstrap results coincide
with the ratio between the two determinations of the lattice
transfer matrix.

\REF\dema{ H.J. de Vega, H. Eichenherr and J.M. Maillet, \NPB{240} (1984) 377
and  \CMP{92} (1984) 507.}
\REF\ddv{C. Destri and H.J. de Vega, \NPB{290} (1987) 363.}
\REF\ddva{C. Destri and H.J. de Vega, \JPA{22} (1989) 1329. }
\REF\leclair{ D Bernard and A LeClair, \CMP{142} (1991) 99.}
\REF\frere{ I. B. Frenkel and N. Yu. Reshetikhin, \CMP{146} (1992) 1.}
\REF\lecsmi{ A. LeClair and F.A. Smirnov, \IMPA{7} (1992) 2997. }
\REF\japs{ B. Davies, O. Foda, M. Jimbo and A. Nakayashiki, \nxl
\CMP{151} 89 (1993)   . }
\REF\zamo{ A.B. Zamolodchikov and Al.B. Zamolodchikov,
                Ann. Phys. {\bf 80} (1979) 253.}
\REF\many{A. LeClair, \PLB{230} (1989) 282. \nxl
          N. Reshetikhin and F. Smirnov, \CMP{131} (1990) 157. }
\REF\jaca{ H.J. de Vega, \NPB ({\sl Proc. Suppl}) {\bf 18 A} (1990) 229. }
\REF\qgba{ C. Destri and H.J. de Vega,
           \NPB{374} (1992) 692 and \nxl \NPB{385} (1992) 361.}
\REF\rev{see for example H.J. de Vega, \IMPA{4} (1989) 2371.}
\REF\eliana{ H.J. de Vega and E. Lopes, \NPB{362} (1991) 261.}
\REF\woyn{ F. Woynarovich, \JPA{15} (1982) 2985.\nxl
           C. Destri and J.H. Lowenstein, \NPB{205} (1982) 369. \nxl
           O. Babelon, H.J. de Vega and C.M. Viallet, \NPB{220} (1983) 13.}
\REF\tba{ Al.B. Zamolodchikov, \NPB{342} (1990) 695 \nxl
          and \NPB{358} (1991) 497. \nxl
          M J Martins, \PLB{240} (1990) 404. \nxl
          T.R. Klassen and E. Melzer, \NPB{338} (1990) 485. }
\REF\leon{ L. A. Takhtadzhyan and L. D. Faddeev, \TMP{21} (1974) 1046. }
\chapter{ Introduction }

When a physical model is integrable it always possess extra conserved
quantities not related to manifest symmetries but presumably with hidden
dynamical symmetries. For 2D lattice models and 2D quantum field theory (QFT),
integrability is a consequence of the Yang-Baxter equation (YBE).

In lattice vertex models, using the $R-$matrix elements to define the local
statistical weights, the monodromy matrix $T_{ab}(\l)$ for a lattice line obeys
the YB algebra:
$$
       R(\l-\mu)\left[T(\l )\otimes
     T(\mu )\right] =\left[T(\mu )\otimes
     T(\l )\right]R(\l-\mu)                     \eqn\nuda
$$
where $\l$ stands for the spectral parameter. Its trace, $t(\l) \equiv
\sum_a T_{aa}(\l) $, provides a commuting family of
operators,
$$
         \left[~t(\l),~t(\mu )~\right]~=~0
$$
for any lattice size. In general the transfer matrix $t(\l)$ is conserved,
since, among all commuting charges, it  generates also the Hamiltonian. It is
now clear from eq.\nuda, that the $T_{ab}(\l)$ do not commute with $t(\l)$ and
are generally not conserved.
That is, for finite lattice size we have only an infinite {\it abelian}
symmetry generated by $t(\l)$ (or by its series expansion coefficients).

In ref.[\dema], a bootstrap construction of  monodromy matrices $\T_{ab}(u)$
was
proposed in a class of integrable QFT. These  $\T_{ab}(u)$ are {\it conserved}
and obey a YB algebra analogous to eq.\nuda.  Hence, this class of QFT enjoy an
infinite YB {\it non--abelian} symmetry generated by the $\T_{ab}(u)$.
(This construction is valid in the infinite space).
A classically conserved limit of  $\T_{ab}(u)$ exists in the class of models
considered in ref. [\dema] where the $R-$matrix is a rational function of $u$.

In the present paper we first show that the bootstrap construction of conserved
$\T_{ab}(u)$ generalizes to integrable models with trigonometric
$R-$matrices such as the sine-Gordon or massive Thirring model.
In such  cases  the classical limit is abelian, as shown explicitly in sec.3.

The main aim of this work is then to investigate and clarify, from a
microscopic point of view, the problem of unveiling the existence
of the infinite YB symmetry of the sG--mT model. In other words, since
lattice models provide regularized version of QFT, we seek an explicit
connection between the lattice and the bootstrap YB algebras.
For this purpose we adopt the so--called light-cone approach, which
is a general method to precisely
derive QFT's as scaling limits of integrable lattice models
[\ddv,\ddva]. One starts from a diagonal-to-diagonal lattice with lines at
angles $2\Th $. The light-cone evolution operators $U_R$ and $U_L$ are
introduced (eq.(5.2)) which define the lattice hamiltonian and momentum
(eq.(5.3)). They
can be expressed in terms of the values at $\l = \pm \Th $ of the
alternating transfer matrix $t(\l,\Th)$ (eqs. (5.5)--(5.6)).
Through algebraic Bethe
Ansatz the ground  state and excited states are constructed in the
thermodynamic limit. When the ground state is  antiferromagnetic, it
corresponds to a Dirac sea of interacting pseudoparticles. Excited states
around it describe particle-like physical excitations. In this way, the sG--mT
model is obtained from the six-vertex model (with anisotropy $\g$)[\ddv].
QFT like multicomponent Thirring models, sigma models and others follow
from various vertex models  [\ddva].

In order to investigate the operators present in such QFT,
it is important to learn how the monodromy operators $T_{ab}(\l, \Th)$ act
on physical states. In the present paper, we explicitly compute
the eigenvalues of the alternating six--vertex transfer matrix
$t(\l,\Th) \equiv \sum_a T_{aa}(\l,\Th)$, on a generic $n-$particle state,
in the thermodynamic limit. The explicit formulae are given by
eqs.(6.32),(6.37) and (6.38). The eigenvalues of $t(\l,\Th)$
turn out to be $i\pi-$periodic and multi--valued functions of $\l$,
each determination of  $t(\l,\Th)$ being a meromorphic function of
$\l$. We call $t^{II}(\l,\Th)$ and $t^{I}(\l,\Th)$ the determinations
associated with the periodicity strips closer to the real axis (see fig. 4).
 The ground--state
contribution $\exp[-iG(\l)_V]$ is exponential on the lattice size, as
expected, whereas the excited states contributions are finite and
express always in terms of hyperbolic functions [see sec. 6].

We then compare these Bethe Ansatz eigenvalues with the eigenvalues of
the bootstrap transfer matrix $\tau(u) \equiv \sum_a \T_{aa}(u)$.
Remarkably enough, we find the following simple relation
between the two results, for $0<\g<\pi/2$ (repulsive regime),
$$
     \tau(u)= t^{II}\bigl({\g\o\pi}u-i{\g\o2},\Th\bigl)
              \,t^I\bigl({\g\o\pi}u-i{\g\o2},\Th\bigl)^{-1}   \eqn\gorda
$$
where $t^{II}(\l,\Th)$ and $t^{I}(\l,\Th)$ have been normalized to one
on the ground state .
 Thus, we succeed in connecting the bootstrap  transfer matrix
$\tau(u)$ of the sG-mT model with the alternating transfer matrix  $t(\l,\Th)$
of the six vertex model. In the thermodynamic limit $\tau(u)$ coincide
with the jump between the two main determinations of $t(\l,\Th)$ . Notice the
renormalization of the rapidity by $ \g/ \pi$ and the precise overall shift
by $i\g/2$ in the argument in order the equality to hold.

We find in addition that $t(\l,\Th)$, for $ 0 < \Im \l < \g/2$, generates
the hamiltonian and momentum
togheter with an infinite number of higher--dimension and higher--spin
conserved abelian charges, through expansion in powers of $e^{\pm\pi\l/\g}$.
We see therefore that the same bare operator generates two kinds of conserved
quantities. Energy and momentum as well the higher--spin abelian charges are
local in the basic fields which interpolate physical particles, whereas the
infinite set of charges obtained from the jump from
 $t^{II}(\l,\Th)$ to $t^{I}(\l,\Th)$
  are nonlocal in the same fields. The
fact that local and nonlocal charges
come from different sides of a natural boundary, clearly
shows that they carry independent information. That is, one cannot
produce the nonlocal charges from the sole knowledge of the local
charges. We also recall that the monodromy matrix $T(\l,\Th)$
can be written in terms of the lattice fermi fields of the mT model
[\ddv], so that local and non local charges do admit explicit expressions in
terms of local field operators.

As a byproduct of this analysis, we find an explicit relation
for the light-cone lattice hamiltonian and momentum  $P_{\pm}$ in terms
of the continuum hamiltonian and momentum $p_{\pm}$ plus an infinite series
of higher conserved continuum charges $I_j^\pm$, playing the r\^ole of
irrelevant
operators,
$$
     P_{\pm}=(P_{\pm})_V + p_{\pm}+{m\o4}\sum_{j=1}^{\infty}
               \left({{ma}\o4}\right)^{2j} I_j^\pm       \eqn\lclhm
$$
where $(P_{\pm})_V$ stands for the ground state contribution.

We expect eqs.\gorda\  - \lclhm, and the discussion in-between, to be
valid for many other integrable models
provided the appropiate rapidity renormalization and imaginary
shift are introduced.

The next natural step after finding the connection \gorda\ between transfer
matrices would be to relate the bare and renormalized monodromies
 $T_{ab}(\l, \Th)$ and  $\T_{ab}(u)$ . This is necessarily more
involved. For $\g \neq 0$ they obey the same YB algebra but
with different anisotropy parameters $\g$ and $\hat \g \equiv {\g \o
{1-\g/\pi}}$, respectively.
The rational case ($\g = 0$) is evidently simpler and it is the only
case where classically conserved monodromies are present.

The quantum monodromy operators $\T_{ab}(u)$ generate a Fock representation
of the $q-$deformed affine Lie algebra $U_q({\hat\G})$
corresponding to the given
$R-$matrix. More precisely, by expanding $\T_{ab}(u)$ in powers of $z=e^u$
around $z=0$ and $z=\infty$, one obtains non--abelian non-local conserved
charges representing the algebra  $U_q({\hat\G})$ on the Fock space of in-- and
out--particles. This connects our approach based on the YB symmetry, to the
$q-$deformed algebraic approach of ref.[\leclair].
$U_q({\hat\G})$ is a Hopf algebra endowed with an universal
$R-$matrix, which reduces to the $R-$ explicitly entering the YB algebra,
upon projection to the finite--dimensional vector space spanned by the
indexes of $\T_{ab}(u)$ [\frere]. In particular, the two expansions around
$z=0$ and $z=\infty$ generate the two Borel subalgebras of  $U_q({\hat\G})$.
A single monodromy matrix  $\T(u)$ is sufficent for this purpose, since this
field--theoretic representation has level zero [\lecsmi]. This fact receives a
new explanation in the light--cone approach, since  $U_q({\hat\G})$
emerges as true symmetry only in the infinite--volume limit above the
antiferromagnetic ground state (with no need to take the continuum limit),
but its action is uniquely defined already on {\it finite} lattices, and
all finite--dimensional representations have level zero.

Besides the conserved operators  $\T_{ab}(u)$, Zamolodchikov-Faddeev
non-conserved operators $Z_\a(\t)$ act by creating particles on physical
states. Their algebra with the $\T_{ab}(u)$ is determined by the two body
S-matrix:
$$
   \T_{ab}(u) Z_\b(\t) = \sum_{c\a}Z_\a(\t)\T_{ac}(u)
                                         S_{b\b}^{c\a}(u+\t) \eqn\ctz
$$
The ZF operators provide a representation of the dynamical symmetry of
$q-$deformed vertex operators in the sense of [\japs].

\chapter{ Bootstrap construction of quantum monodromy operators.}

We briefly review in this section the work of refs. [\dema ] where the
exact (renormalized) matrix elements of a quantum monodromy matrix
$\T_{ab}(u)$ ($u$ is the generally complex spectral parameter) were derived
using a bootstrap--like approach for a class of integrable local QFT's. In such
theories there is no particle production  and the  $S-$matrix factorizes. The
two--body $S-$matrix then satisfies the Yang--Baxter (YB) equations. Moreover,
in the models considered in  refs.[\dema ] (the O(N) nonlinear sigma
model, the SU(N) Thirring model and the 0(2N) Gross--Neveu model), thanks to
scale invariance there exist classically conserved monodromy matrices. In
general, the quantum $\T_{ab}(u)$ can be constructed by fixing
its action on the Fock space
of physical in and out many--particle states. The starting point are the
following three general principles:

\item{a)}
$\T_{ab}(u)$, $a,b=1,2,\ldots,n$, exist as quantum operators and are conserved.
\item{b)}
$\T_{ab}(u)$ fulfil a quantum factorization principle.
\item{c)}
$\T_{ab}(u)$ is invariant under P, T and the internal symmetries of the theory.

The quantum factorization principle  referred above under b)
is nowadays called the "coproduct rule". This means that there exists
the following relation between the action of $\T_{ab}(u)$
on $k-$particles states and its action on one--particle states
$$
   \T_{ab}(u)\ket{\t_1\a_1,\t_2\a_2,\ldots,\t_k\a_k}_{in}=
   \sum_{a_1a_2\ldots a_{k-1}} \! \T_{aa_1}(u)\ket{\t_1\a_1}
   \T_{a_1a_2}(u)\ket{\t_2\a_2} \dots \T_{a_{k-1}b}(u)\ket{\t_k\a_k} \eqn\tin
$$ $$
   \T_{ab}(u)\ket{\t_1\a_1,\t_2\a_2,\ldots,\t_k\a_k}_{out}=
   \sum_{a_1a_2\ldots a_{k-1}} \!\! \T_{a_1b}(u)\ket{\t_1\a_1}
   \T_{a_2a_1}(u)\ket{\t_2\a_2} \dots \T_{aa_{k-1}}(u)\ket{\t_k\a_k} \eqn\tout
$$
where $\t_j$ and $\a_j \; (1\le j\le k)$ label the rapidities and
the internal quantum numbers of the particles, respectively, in the asymptotic
in and out states. Hence it is understood that $\t_i>\t_j$ for $i>j$.

Although $\T_{ab}(u)$ acts differently on in and out states, the assumption
of conservation is nonetheless consistent. All the eigenvalues of a
maximal commuting subset of $\{\T_{ab}(u),\;a,b=1,2,\ldots,n,\; u\in\CC\}$
are identical for in and out states with given rapidities. Indeed the two in
and out forms of the action on the internal quantum numbers are related by the
unitary permutation
$\ket{\a_1,\a_2,\ldots,\a_k}\to \ket{\a_k,\a_{k-1},\ldots,\a_1}$.

Furthermore, principles (a) and (c) imply that $\T_{ab}(u)$ acts in a trivial
way on the physical vacuum state $\ket{0}$:
$$
           \T_{ab}(u)\ket{0}=\delta_{ab}\ket{0}                  \eqn\onvac
$$
This also fixes the normalization of $\T_{ab}(u)$ in agreement with the
classical limit [\dema].

An immediate consequence of point (b) is that when $\T_{ab}(u)$ is expanded
in powers of the spectral parameter $u$, it generates an infinite set of
noncommuting and  nonlocal conserved charges. This is the clue to the matching
of the quantum monodromy matrix with its classical counterpart which is written
nonlocally in terms of the local fields.

The main result in refs.[\dema ] was to derive from (a), (b) and (c) the
explicit matrix elements of $\T_{ab}(u)$ on one--particle states.
This result can be written as
$$
           \bra{\t\a}\T_{ab}(u)\ket{\t^\prime \b}= \delta(\t-\t^\prime)
           S^{a\a}_{b\b}(\kappa(u)+\t)                     \eqn\matel
$$
where $S^{a\a}_{b\b}(\t-\t^\prime)$ stands for the $S-$matrix of two--body
scattering
$$
       \ket{\t b,\t^\prime \b}_{in}=\sum_{a\a}\ket{\t a,\t^\prime \a}_{out}
               S^{a\a}_{b\b}(\t-\t^\prime)                    \eqn\smat
$$
and $\kappa(u)$ is an odd function of $u$. Notice that this requires
the presence in the model of particles with indices $a, b, ...$
as internal state labels. In the simplest situation these new labels
coincide with those of the original particles. The appearance of a nontrivial
``renormalization" $u \to \kappa(u)$ is to be expected when there exist
a definition of the spectral parameter outside the bootstrap itself. This is
the case of the models of refs.[\dema ], which posses Lax pairs and
auxiliary problems which fix the definition of $u$. Here we shall
adopt the purely bootstrap viewpoint and fix the definition of $u$ so
that $\kappa(u)=u$.
In principle, an extra $u-$ and $\t-$dependent phase factor may
appear in the r.h.s. of eq.\matel.
However, no phase showed up in the specific models of refs.[\dema ],
when nonperturbative checks were performed using the operator product
expansion.
Eq.\matel\ can be written in a more suggestive way as
$$
          \T_{ab}(u)\ket{\t\b}= \sum_\a\ket{\t\a}
            S^{a\a}_{b\b}(u+\t)                      \eqn\opa
$$
This equation, when combined with eqs.\tin\ and \tout, completely defines
the quantum monodromy operators in the Fock space. From the YB equations
satisfied by the $S-$matrix it then follows that $\T_{ab}(u)$ fulfils the
YB algebra
$$
        {\hat R}(u-v)\left[\T(u)\otimes \T(v)\right]=
        \left[\T(u)\otimes \T(v)\right]{\hat R}(u-v)                \eqn\yba
$$
where ${\hat R}^{a\a}_{b\b}(u)=S^{\a a}_{b\b}(u)$.
It should be stressed that the conservation of $\T_{ab}(u)$ implies that
this YB algebra is a true {\it non--abelian infinite symmetry algebra} of the
relativistic local QFT. On the contrary the r\^ole of the YB
algebra in integrable vertex and face models on finite lattices or in
nonrelativistic quantum models is that of a dynamical symmetry
underlying the Quantum Inverse Scattering Method. In these latter cases,
only the transfer matrix, namely
$$
          \tau(u)=\sum_a \T_{aa}(u)                            \eqn\trans
$$
is conserved. Since $\left[\tau(u),\tau(v)\right]=0$, the transfer matrix
just generates an abelian symmetry.

The dynamical symmetry underlying the integrable QFT includes in addition
non--conserved operators $Z_\a(\t)$ which create the particle eigenstates out
of the vacuum. In the bootstrap framework they can be introduced \`a la
Zamolodchikov--Faddeev, by setting
$$\eqalign{
   &\ket{\t_1\a_1,\t_2\a_2,\ldots,\t_k\a_k}_{in}  =
   Z_{\a_k}(\t_k)  Z_{\a_{k-1}}(\t_{k-1}) \ldots  Z_{\a_1}(\t_1) \ket0 \cr
  &\ket{\t_1\a_1,\t_2\a_2,\ldots,\t_k\a_k}_{out}  =
   Z_{\a_1}(\t_1)  Z_{\a_2}(\t_2) \ldots  Z_{\a_k}(\t_k) \ket0 \cr} \eqn\zf
$$
with the fundamental commutation rules
$$
     Z_{\a_2}(\t_2)  Z_{\a_1}(\t_1) =\sum_{\b_1\b_2}
     S_{\a_1\a_2}^{\b_1\b_2}(\t_1-\t_2)  Z_{\b_1}(\t_1)  Z_{\b_2}(\t_2)
\eqn\zzf
$$
Combining now eqs. \tin, \tout, \opa\ and \zzf, we obtain the algebraic
relation between monodromy and Zamolodchikov--Faddeev operators:
$$
   \T_{ab}(u) Z_\b(\t) = \sum_{c\a}Z_\a(\t)\T_{ac}(u)
                                         S_{b\b}^{c\a}(u+\t) \eqn\fcr
$$
Together with eqs. \yba\ and \zzf, these relations close the complete dynamical
algebra of an integrable QFT.
For the XXZ spin chain in the regime  $|q| < 1$, the ZF operators have been
identified in ref. [\japs] with special vertex operators (or representation
intertwiners of the relevant $q-$deformed affine Lie algebra). They are
uniquely characterized by being solutions of the $q-$deformed
Knizhnik--Zamolodchikov equation and by their normalization [\frere].

\chapter{ Yang-Baxter symmetry in the sine-Gordon model}

In refs.[\dema ]  the infinite YB symmetry was explicitely
considered and exhibited for classically scale--invariant models
like the O(N) nonlinear sigma, the SU(N) Thirring and the 0(2N) Gross--Neveu
models. Indeed, it is this scale--invariance which guarantees the
conservation also of the nondiagonal elements of the classical monodromy
matrix. An important example of
integrable QFT  which does not belong to this category is the sine--Gordon (sG)
model. The presence of mass at the classical level implies that only its
transfer matrix is conserved. Then the classical symmetry algebra is commuting
(as Poisson brakets) and admits a basis  of  {\it local} conserved charges. It
is well known that these charges survive quantization, leading to factorization
of the scattering  with corresponding YB equations satisfied by the two--body
$S-$matrix [\zamo ].

It appears therefore natural to apply the general bootstrap
methods of the previous section also to the sG model. The quantum $\T_{ab}(u)$
(with $a,b=\pm$)
defined in this way is assumed to be conserved from the outset, and does
not reduce to its classical counterpart in the classical limit.
Let us consider first soliton and antisoliton states. They
have mass $M$ and a charge $\a = +1\, (-1)$ for solitons (antisolitons).
Let us recall that the solitons (antisolitons) are the fermions
(antifermions) of the massive Thirring model, which is equivalent to the
sG model as QFT in 2D Minkowski space.

Eq.\opa\ in the one-particle soliton/antisoliton sector
takes now the form
$$\eqalign{
        \T_{\pm\pm}(u)\ket{\t\pm} &= S(u+\t)\ket{\t\pm}   \cr
        \T_{\pm\pm}(u)\ket{\t\mp} &= S_T(u+\t)\ket{\t\mp} \cr
        \T_{\pm\lowmp}(u)\ket{\t\pm} &= S_R(u+\t)\ket{\t\mp} \cr
        \T_{\pm\lowmp}(u)\ket{\t\mp} &= 0                        \cr}  \eqn\sg
$$
where $S(\t)$ is the soliton/soliton scattering amplitude, while $S_T(\t)$ and
$S_R(\t)$ are, respectively the transmission and reflection amplitudes of the
soliton/antisoliton scattering. Explicitly they read [ \zamo\ ]:
$$\eqalign{
    S(\t) &=\exp\,i\int_0^\infty {dk\o k}
            {{\sinh(\pi/{\hat\g}-1)k/2}\o{\sinh(\pi k/2{\hat \g})}}
            {{\sin k\t/\pi}\o{\cosh k/2}}                 \cr
   S_T(\t) &=S(i\pi-\t) \cr
   S_R(\t) &= S(\t)
            {\sin{\hat\g}\o{\sin\left[{\hat\g}(1-\t/i\pi)\right]}}\cr} \eqn\amp
$$
where ${\hat\g}$ is related to the usual sG coupling constant $\b$ by
$$
       {{\hat\g}\o \pi}= {{8\pi}\o{\b^2}}-1                    \eqn\gambeta
$$
The action of $\T_{ab}(u)$ on multiparticle soliton/antisoliton states
is obtained  by simply inserting eqs. \sg\ into eqs.\tin\ and \tout.
Notice that  $S(\t)$, $S_T(\t)$ and $S_R(\t)$ posses essential
singularities at $\b^2 = 0$. That is,
$$
S(\t)\buildrel{\b\to0}\over =\exp\,-i{{16\pi}\o{\b^2}}\int_0^\infty {dk\o k^2}
            {\tanh(k/2)}{\sin k\t/\pi} \equiv S_c(\t)    \eqn\singu
$$
Hence the quantum monodromy matrix is singular in the free boson limit $\b=0$.
Of course it is regular, although trivial ($\T_{ab}(u)=\delta_{ab}$),
in the free fermion limit $\hat\g=\pi$. In the {\it classical} limit instead,
we must replace the scattering amplitudes with the analogous quantities
computed for soliton field configurations in the classical sG model, namely
$$
      S(\t)=S_T(i\pi-\t)= S_c(\t) \;,\quad S_R(\t)=0
$$
(there is no soliton reflection at the classical level). Hence $\T_{ab}(u)$
becomes diagonal and the YB algebra becomes abelian. This is consistent with
the fact that the integrals of motion of the classical sG equation are all
in involution.

Besides solitons and antisolitons states, the sG-MTM model
possess breathers states for ${\hat\g}>\pi$. These particles are labeled
by and index $n$ running from 1 to $[{\hat\g}/\pi]-1$ (where [..]
stands for integer part) and have  masses
$$
m_n = 2M \sin ( {{n \pi^2} \o {2 \hat\g}})
               \eqn\espma
$$
[$n = 1$ corresponds to the fundamental particle associated to the
sG-field ].
Applying the general formula \matel\ to these particle states
yields
$$
\bra{\t n}\T_{ab}(u)\ket{\t^\prime m}= \delta(\t-\t^\prime)
          \delta_{nm} \delta_{ab} S_{n}(u-\t)
\eqn\masego
$$
 where $\ket{\t n}$ stands for a $n$th. breather state with rapidity
$\t$ and $S_{n}(\t-\t^\prime)$ is the soliton- breather S-matrix. That
is [ \zamo\ ] ,
$$
  S_{n}(\t)= {{ \sinh\t + i \cos{n\pi^2 \o {2\hat\g}} }\o
 { \sinh\t - i \cos{n\pi^2 \o {2\hat\g}} }}
\prod_{l=1}^{n-1}{ {\sin^2( \left[ {n \o 2}-l
\right]{\pi^2 \o {2\hat\g}} - {\pi \o 4} + {{i \t} \o 2})}\o
{\sin^2( \left[ {n \o 2}-l
\right]{\pi^2 \o {2\hat\g}} - {\pi \o 4} - {{i \t} \o 2})}} \quad
      \eqn\sbsm
$$
In particular,
$$
       S_{1}(\t) = {{ \sinh\t + i \cos{\pi^2 \o {2\hat\g}} }\o
        { \sinh\t - i \cos{\pi^2 \o {2\hat\g}} }}                  \eqn\pfsg
$$
We conclude that $\T_{ab}(u)$ has a rather trivial action on
breather states
$$
          \T_{ab}(u)=\delta_{ab}S_B(u)                           \eqn\matbr
$$
where $S_{B}(u)$ is a diagonal operator with eigenvalues $S_n(u-\t) $
on the $n$th breather state.
In conclusion, we have uncovered the infinite YB symmetry of
the sG-mT model providing the explicit form of its conserved
operators on all the asymptotic states.

It is instructive to study the $u \to \infty$ limit of the YB
operator $\T_{ab}(u)$. We find from eqs. \sg-\amp\  for $u \to \infty$ ,
$$
     \T_{ab}(u) = \exp({ia{{4 \pi^2}\o{\b^2}}\sigma_3})\; \delta_{a b}+
      2i e^{4i {\pi^2\o{\b^2}}} \exp(-{{8\pi u}\o{\b^2}}) Q_{b} ( 1 -
      \delta_{a b}) + O(e^{-{{16\pi u}\o{\b^2}}})               \eqn\asit
$$
Here $\b$ is the usual sine-Gordon coupling constant and
$Q_a$ acts on one-particle soliton/antisoliton states as
$$
               Q_{a}= e^{-{{\g\t}\o\pi}}\sigma_a                   \eqn\qquu
$$
This is a $SU(2)_q$ generator for the spin 1/2
representation. (For spin 1/2, SU(2) and $SU(2)_q$ generators
coincide). Using eq.\asit\ and the coproduct relations \tin\
and \tout\ , we find that eq.\asit\ holds as it stands on two
(or more) particle states but now with
$$
\s_3 = \s_3^{(1)} + \s_3^{(2)}  \qquad  , \qquad
  Q_{a} = e^{-{ia{{4 \pi^2}\o{\b^2}}}\sigma_3^{(1)}}\,Q_{a}^{(2)}+
  Q_{a}^{(1)}\, e^{{ia{{4 \pi^2}\o{\b^2}}}\sigma_3^{(2)}}
                        \eqn\copro
$$
Analogous relations hold for multiparticle states.
This tells us that $Q_{a}$ and $\s_3 $ are related to $SU(2)_q$ generators
with
$$
q = e^{{8i \pi^2}\o{\b^2}}
                \eqn\valoq
$$
as \vskip -0.5 true cm
$$
J_+ =Q_+ \quad , \quad J_- =Q_-^{\dagger} \quad , \quad J_z = \s_3
                  \eqn\repre
$$
Alternatively, we can make the identification
$ q =e^{-{{8i \pi^2}\o{\b^2}}} $ with:
$$
J_+ =Q_+^{\dagger} \quad , \quad J_- =Q_- \quad , \quad J_z = \s_3
                  \eqn\reprep
$$
A nonlocal charge equivalent to $Q_a$ studied in
ref.[\many].
The fact that YB generators for $ u = \infty$  yield $SU(2)_q$
generators in this way is typical of periodic boundary
conditions [\jaca ]. For fixed boundary conditions (that is
scattering of particles between two walls) the connection is
much cleaner [\qgba ].

\chapter{ Bethe Ansatz at the bootstrap level }

The maximal abelian subalgebra of the YB algebra \yba\ is generated by the
transfer matrix $\tau(u)$ (eq.\trans). With respect to this subalgebra, the
remaining elements of $\T_{ab}(u)$ act as generalized raising and lowering
operators. This observation provides the basis for the so--called Algebraic
Bethe Ansatz, which is a purely algebraic method to construct the eigenvectors
and the eigenvalues of $\tau(u)$ [\rev ]. The crucial starting point is the
identification of the highest weight states annihilated by the raising
operators. Since particles are conserved in an integrable QFT model, one can
restrict the problem to states with a fixed number, say $k$, of particles.

In the case of the sG model the
highest weight states are the ferromagnetic states containing only solitons,
that is the states $\ket{\t_1+,\t_2+,\ldots,\t_k+}$ with the highest
possible value $J_z=k/2$ of the $z-$projection of the $SU(2)_q$
spin in the sector with $k$ particles. On such states the monodromy
matrix $\T_{ab}(u)$ is indeed upper triangular (compare eqs. \tin\ and \tout\
with eqs. \sg).
The rapidities $\t_n$ of the solitons are
arbitrary and act as fixed parameters in the problem, since they are left
unchanged by the action of $\T_{ab}(u)$.
Then the BA in--eigenstates of $\tau(u)=\T_{++}(u)+\T_{--}(u)$
with $k-m$ solitons and $m$ antisolitons can be written
$$
   \B(u_1)\B(u_2)\ldots \B(u_m)
            \ket{\t_1+,\t_2+,\ldots,\t_k+}_{in}
$$
where $\B(u) \equiv \T_{+-}(u+i\pi/2)$ act as lowering operators of $J_z$
and the distinct numbers $u_1,u_2,\ldots,u_m$ must satisfy
the BA equations
$$
     \prod_{n=1}^k
           {{\sinh {\hat\g} [i/2+(u_j+\t_n)/\pi]} \o
            {\sinh {\hat\g} [i/2-(u_j+\t_n)/\pi]}} =
     -\prod_{r=1}^m
           {{\sinh {\hat\g} [+i+(u_j-u_r)/\pi]}  \o
            {\sinh {\hat\g} [-i+(u_j-u_r)/\pi]}}       \eqn\hbae
$$
The eigenvalues $\xi(u)$ of $\tau(u)$ on the BA states read
$$\eqalign{
     \xi(u)   &= \left\{ \prod_{n=1}^k S(u+\t_n) \right\}
                 \left[\, \xi_+(u) + \xi_-(u) \right]   \cr
     \xi_+(u) &=  \prod_{j=1}^m
                   {{\sinh {\hat\g} [i/2+(u-u_j)/\pi]}  \o
                    {\sinh {\hat\g} [i/2-(u-u_j)/\pi]}}  \cr
     \xi_-(u) &= \left[\prod_{n=1}^k
                       {{\sinh {\hat\g} (u+\t_n)/\pi} \o
                        {\sinh {\hat\g} [i-(u+\t_n)/\pi]}} \right]
                  \prod_{j=1}^m
                   {{\sinh {\hat\g} [3i/2+(u-u_j)/\pi]}  \o
                    {\sinh {\hat\g} [-i/2+(u-u_j)/\pi]}}   \cr}     \eqn\heig
$$
Up to the common factor $\prod_{n=1}^k S(u+\t_n)$, $\xi_\pm(u)$
is just the contribution coming from $\T_{\pm\pm}(u)$.
It is clear, moreover, that the presence of breathers introduce no
further complications, due to the diagonal action \matbr\ of the
monodromy matrix on breather states.

Eqs. \hbae\ and \heig\ follow directly from the YB algebra \yba\ satisfied
by construction by the bootstrap monodromy matrix. This algebraic Bethe
Ansatz can be generalized to a whole class of integrable field
theories where the bootstrap construction of sec. 2 applies.
Furthermore, let us observe that the diagonalization of the bootstrap transfer
matrix represents the basic step in the so--called Thermodynamic BA,
which is a way to obtain off--shell exact results on the integrable
relativistic QFT at hand. In fact, the transfer matrix $\tau(u)$, as
trace of the monodromy matrix (eq.\trans), is directly related to the
multiscattering amplitudes suffered by each particle in a system of $k$
particles confined on a ring, namely
$$
      \tau(-\t_j)=S_{jk} \ldots S_{j\,j+1}S_{j\,j-1} \ldots S_{j1}  \eqn\multis
$$
where $j=1,2,\ldots,k$ and the two--body matrices $S_{ij}$ are defined by
$$
    S_{ij} \ket{\t_1\a_1,\ldots,\t_k\a_k} = \sum_{\b_i\b_j}
           \biggl(\prod_{n \ne i,j}\delta^{\b_n}_{\a_n} \biggr)
  S^{\a_j\a_k}_{\b_i\b_j}(\t_i-\t_j)\ket{\t_1\b_1,\ldots,\t_k\b_k}
$$
By periodicity, eqs. \multis\ and \heig\ determine the quantization of
the momentum of each particle in the standard way
$$
        \xi(-\t_j)\exp\left(i\, m_j L\sinh\t_j \right) = 1          \eqn\quant
$$
where $L$ is the length of the ring. Together with the BA equations
\hbae\ for the roots $u_1,u_2,\ldots,u_m$ (the so--called {\it magnon}
parameters), these new equations provide the basis for the TBA [\tba].

\chapter{ Light-cone lattice regularization.}

In order to obtain a first--principles, microscopic understanding of the
bootstrap picture presented above, we now consider the
integrability--preserving lattice regularization of an integrable relativistic
QFT defined by the so--called light-cone approach [\ddv,\ddva] to vertex
models.

In this approach one starts from the discretized Minkowski 2D space--time
formed by a regular diagonal lattice of right--oriented and
left--oriented straight lines (see fig. 1). These represent true world--lines
of  ``bare'' objects (pseudo--particles) which are thus naturally divided in
left-- and right--movers. The right--movers have all the same positive rapidity
$\Th$, while the left--movers have rapidity $-\Th$. One can regard $\Th$ as a
cut--off rapidity, which will be appropriately taken to infinity in the
continuum limit. Furthermore, we shall denote by $\V$ the Hilbert space of
states of a pseudo--particle (we restrict here to the case in which $\V$ is the
same for both left-- and right--movers and has finite dimension $n$, although
more general situations can be considered).

The dynamics of the model is fixed by the microscopic transition amplitudes
attached to each intersection of a left-- and a right--mover, that is to each
vertex of the lattice. This amplitudes can be collected into linear
operators $R_{ij}$, the local $R-$matrices, acting non--trivially only
on the space $\V_i\otimes\V_j$ of $i$th and $j$th pseudo--particles.
$R_{ij}$ thus represent the relativistic scatterings of left--movers on
right--movers and depend on the rapidity difference $\Th-(-\Th)=2\Th$,
which is constant throughout the lattice. Moreover,
by space--time translation invariance any other parametric dependence
of $R_{ij}$ must be the
same for all vertices. We see therefore that attached to each vertex
there is a matrix $R(2\Th)^{cd}_{ab}$, where $a,b,c,d$ are labels for
the states of the pseudo--particles on the four links stemming out of
the vertex, and take therefore $n$ distinct values (see fig. 1). This is the
general framework of a  vertex model.
The difference with the standard statistical interpretation is
that the Boltzmann weights are in general complex, since we should
require the unitarity of the matrix $R$. In any case,
the integrability of the model is guaranteed whenever $R(\l)^{cd}_{ab}$
satisfy the Yang--Baxter equations
$$
        R_{ij}(\l) R_{jk}(\l+\mu) R_{ij}(\mu) =
        R_{jk}(\mu) R_{ij}(\l+\mu) R_{jk}(\l)               \eqn\ybeqs
$$
For periodic boundary conditions, the one--step
light--cone evolution operators $U_L(\Th)$ and $U_R(\Th)$, which act on the
''bare'' space of states $\H_N=(\otimes\V)^{2N}$ , ($N$ is the number of
sites on a row of the lattice, that is the number of diagonal lines),
are built from the local $R-$matrices $R_{ij}$ as [\ddv]
$$\eqalign{
         U_R(\Th)&= U(\Th)V \;,  \qquad  U_L(\Th) =U(\Th)V^{-1} \cr
         U(\Th)&=R_{12}R_{34}\ldots \sub R{2N-1\,2N}  \cr}  \eqn\evol
$$
where $V$ is the one-step space translation to the right.
$U_R$ ( $U_L$ ) evolves states by one step in right (left) light--cone
direction. $U_R$ and $U_L$ commute and their product $U=U_R U_L$ is the unit
time evolution operator. The graphical representation of $U$ is given by the
section of the diagonal lattice with fat lines in fig. 1.
If $a$ stands for the lattice spacing, the lattice hamiltonian $H$ and total
momentum $P$ are naturally defined through
$$
             U=e^{-iaH} \;,\qquad  U_R U_L^{-1}=e^{iaP}    \eqn\evolu
$$
The action of other fundamental operators is naturally defined on the same
Hilbert space $\H_N$. These are the $n^2$ Yang-Baxter operators for $2N$ sites,
which are conventionally grouped into the $n\times n$ monodromy matrix
$T(\l)=\{T_{ab}(\l),\;a,b=1,\ldots,n\}$.
One usually regards the indices $a,b$ of $T_{ab}$ as {\it horizontal}
indices fixing the out-- and in--states of a reference pseudo--particle.
Then $T(\l)$ is defined as horizontal coproduct of order $2N$ of the
local vertex operators $L_j(\l)=R_{0j}(\l)P_{0j}$, where $0$ label the
reference space and $P_{ij}$ is the transposition in $\V_i\otimes\V_j$ .
Explicitly
$$
        T(\l) = L_1(\l)L_2(\l) \ldots L_{2N}(\l)
$$
The inhomogenuous generalization $T(\l,{\vec\om\,})$ then reads
$$
  T(\l,{\vec\om\,}) = L_1(\l+\om_1)L_2(\l+\om_2) \ldots L_{2N}(\l+\om_{2N})
$$
and has the graphical representation of fig. 2.
The formal structure of this expression is identical to that of eq.\tin.
In fact $L_j(\l+\om_j)$ can be regarded as the scattering of the $j$th
pseudo--particle carrying formal rapidity $\om_j$ with the reference
pseudo--particle carrying formal rapidity $-\l$. In the same way,
thanks to eq.\opa, the single particle terms in eq.\tin\ represent
the scattering of the corresponding particle on a reference particle
carrying physical rapidity $-u$. In the case of our diagonal lattice
of right-- and left--moving pseudoparticles, there exists a specific,
physically relevant choice of the inhomogeneities, namely
$$
   \om_k=(-1)^k \Th \;,\quad k=1,2,\ldots 2N       \eqn\alt
$$
leading to the definition of the {\it alternating} monodromy matrix
$$
      T(\l,\Th)\equiv T(\l,\{\om_k=(-1)^k\Th\})                    \eqn\altmon
$$
In fact, the evolution operators $U_L(\Th)$ and $U_R(\Th)$ can be
expressed in terms of the alternating transfer matrix
$t(\l,\Th)={\rm tr}_0 T(\l,\Th)$ as [\ddva]
$$
      U_R(\Th)= t(\Th,\Th) \;,\quad U_L(\Th)= t(-\Th,\Th)^{-1}   \eqn\ttou
$$
At any rate, no matter how the $\om_k$ are chosen, the monodromy matrix
$T(\l,{\vec\om\,})$ fulfils the YB algebra
$$
       R(\l-\mu)\left[T(\l,{\vec\om\,})\otimes
     T(\mu,{\vec\om\,})\right] =\left[T(\mu,{\vec\om\,})\otimes
     T(\l,{\vec\om\,})\right]R(\l-\mu)                     \eqn\bareyba
$$
just as the quantum $\T(u)$ satisfies the YB algebra \yba. We see that
the ``bare'' YB algebra involves the finite--dimensional operators
$T_{ab}(\l,{\vec\om\,})$ and, correspondingly, the ``bare'' $R-$matrix
$R(\l)$ defines it.

Notice that $T(\l,\Th)$ fails to be conserved on the lattice only
because of boundary effects. Indeed from fig. 3, which graphically represents
the insertion of  $T(\l,\Th)$ in the lattice time evolution, one readily sees
that $U$ and  $T(\l,\Th)$ fail to commute only because of the free ends of the
horizontal line. For all vertices in the bulk, the graphical interpretation
of the YB equations \ybeqs, namely that lines can be freely pulled through
vertices, allows to move  $T(\l,\Th)$ up or down, that is to freely commute it
with the time evolution. The problem lays at the boundary: if periodic boundary
conditios are assumed, then the free horizontal ends of  $T(\l,\Th)$ cannot
be dragged along with the bulk, unless they are tied up, to form the
transfer matrix  $t(\l,\Th)$. After all, for p.b.c., the boundary is
actually equivalent to any point of the bulk and thus  $t(\l,\Th)$ commutes
with $U$, as obvious also from eqs. \ttou\ and the general fact that
$[t(\l,\Th),\,t(\mu,\Th)]=0$. One might think that the thermodynamic limit
$N\to\infty$, by removing infinitely far away the troublesome free ends of
$T(\l,\Th)$, will allow for its conservation and thus for the existence
of an exact YB symmetry with bare $R-$matrix. The situation however is not so
simple: first of all one must fix the Fock sector of the $N\to\infty$
non--separable Hilbert space in which to take the thermodynamic limit.
Different choices leads to different phases with dramatically different
dynamics. Then the non--local structure of  $T(\l,\Th)$ must be taken into
account. It is evident, for instance, that in the spin--wave Fock sector above
ferromagnetic reference states  $T(\l,\Th)$ can never be conserved. Indeed, the
working  itself of the Quantum Inverse Scattering Method, where energy
eigenstates are built applyind non--diagonal elements of $T(\l,\Th)$ on
a specific ferromagnetic reference state, of course
depends on $T(\l,\Th)$  {\it not} commuting with the hamiltonian!

 From the field--theoretic point of view, the most interesting phase is the
antiferromagnetic one, in which the ground state plays the r\^ole of densily
filled {\it interacting} Dirac sea (this holds for all known integrable lattice
vertex models [\ddv,\ddva,\rev]).
The corresponding Fock sector is formed by particle--like
excitations which become relativistic massive particles within the scaling
limit proper of the light--cone approach [\ddva]. This consists in letting
$a\to 0$ and $\Th\to\infty$ in such a way that the physical mass scale
$$
                \mu=a^{-1} e^{-\kappa\Th}                  \eqn\mass
$$
stays fixed. Here $\kappa$ is a model--dependent parameter which
for the so--called {\it rational} class of integrable model (to this class
belong the models considered in ref. [\dema]) takes the general form [\eliana]
$$
            \kappa={{2\pi\, t}\o{h\, s}}                    \eqn\kconst
$$
where $h$ is the dual Coxeter number of the underlying Lie algebra, $s$
equals 1, 2 or 3 for simply, doubly and triply laced algebras,
respectively, and  $t=1 \;(t=2)$  for non--twisted (twisted) algebras.
For the class of model characterized by a trigonometric $R-$matrix
(with anisotropy parameter $\g$) the expression $\kconst$ for $\kappa$
is to be divided by $\g$ [\eliana].

The ground state or (physical vacuum) and the particle--like excitations of
this antiferromagnetic phase are extremely more complicated than those of the
ferromagnetic phase. It is therefore very hard to control, in the limit
$N\to\infty$, the action of the
alternating monodromy matrix $T(\l,\Th)$ on the particle--like BA
eigenstates of the alternating transfer matrix  $t(\l,\Th)$. On the other
hand, since these particles enjoy a factorized scattering, one can proceed
according to the general tenets of the bootstrap approach described in sec. 2.
In this way one constructs the bootstrap monodromy matrix $\T(u)$ and it is
natural to search for an explicit connection between  $\T(u)$ and $T(\l,\Th)$.
It is a connection like this that would provide the microscopic interpretation
of the bootstrap results.

In order to study the infinite volume limit of $T(\l,\Th)$ on the
physical Fock space (that is, finite energy excitations around the
antiferromagnetic vacuum), one needs to compute scalar products of
Bethe Ansatz states to derive relations like \matel\ or \opa\ with
 $T(\l,\Th)$ instead of $\T(\l,\Th)$ in the l.h.s. Since this kind of
calculations are indeed possible but rather involved, we start by
computing the eigenvalues of  $t(\l,\Th)$ on a generic state of the
physical Fock space. Then, we shall compare these eigenvalues with
those of  $\tau(u)$. This will tell us whether the bare and the
renormalized YB algebras have a common abelian subalgebra.
Notice that this fact alone would provide a microscopic basis
for the TBA, which originally relies solely on the bootstrap.

We shall consider once more the sG model as example, although the same
result would apply to any integrable QFT admitting a light--cone
lattice regularization. This class of models contains also the
O(N) nonlinear sigma model and the $SU(N)$ Thirring model considered
from the bootstrap viewpoint in refs.[\dema ].

The integrable light--cone lattice regularization of
the sG--mT model is provided the six-vertex model [\ddv].
Therefore, the space $\V$ is two--dimensional and the unitarized
local $R-$matrices can be written
$$\eqalign{
       R_{jk}(\l) =&{{1+c}\o2}+{{1-c}\o2}\s^z_j\s^z_k
                     +{b\o2}(\s^x_j\s^x_k + \s^y_j\s^y_k)      \cr
             b =&{{\sinh\l}\o{\sinh(i\g-\l)}}\;,\quad
             c ={{\sinh i\g}\o{\sinh(i\g-\l)}}       \cr}       \eqn\rma
$$
where $\g$ is commonly known as anisotropy parameter.

The standard Algebrized BA can be applied to the diagonalization of
the alternating transfer matrix $t(\l,\Th)$ with the following
results [\ddv,\ddva,\jaca].
The BA states are written
$$
          \Psi ( {\vec\l}\,) = B(\l_1) .... B(\l_M ) \Omega            \eqn\aba
$$
where ${\vec\l}\equiv (\l_1,\l_2,\ldots,\l_M)$,
$B(\l_i) = T_{+-}(\l_i + i\g/2 ,\Th)$  and $\Omega$ is the
ferromagnetic ground-state (all spins up).
They are eigenvectors of  $t(\l,\Th)$
$$
t(\l,\Th) \Psi ( {\vec\l}\, ) =
 \Lambda (\l ; {\vec\l}\, )
\Psi ( {\vec\l}\, )
                         \eqn\eigeq
$$
provided the $ \l_i $ are all distinct roots of the ``bare'' BA equations
$$
  \left({{\sinh[i\g /2+ \l_j - \Th]}\o{\sinh[i \g /2-\l_j + \Th]} }
     \; {{\sinh[i\g /2+ \l_j + \Th]}\o{\sinh[i \g /2-\l_j - \Th]} }\right)^N =
     - \prod_{k=1}^M
     {{\sinh[+i\g +\l_j - \l_k]}\o{\sinh[-i \g +\l_j -\l_k] }}
                             \eqn\bae
$$
The eigenvalues $ \Lambda (\l ; {\vec\l}\, )$ are the sum of a
contribution coming from
$A(\l)=T_{++}(\l,\Th)$ and one coming from $D(\l)=T_{--}(\l,\Th)$,
$$
\Lambda (\l ; {\vec\l}\, ) = \Lambda_A(\l ; {\vec\l}\, )
+ \Lambda_D(\l ; {\vec\l}\, )                    \eqn\eigev
$$
Here
$$\eqalign{
    \Lambda_A(\l ; {\vec\l}\, ) &=
      \exp\left[ -i G(\l , {\vec\l}\, )\right] \cr
    \Lambda_D(\l ; {\vec\l}\, ) &=
        e^{-iN \left[ \phi(\l-i\g/2 - \Th , \g /2) +
          \phi(\l-i\g/2 + \Th, \g /2) \right]}
          \exp{[iG(\l - i\g ,  {\vec\l}\, )]} \cr}    \eqn\llamm
$$
and
$$
G(\l,{\vec\l}\,) \equiv \sum_{j=1}^M \phi(\l - \l_j, \g/2) \;,\quad
 \phi(\l,\g) \equiv i\log {  {\sinh(i\g+\l)}\o{\sinh(i\g-\l)}}     \eqn\gee
$$
$G(\l,{\vec\l}\,)$ is manifestly a periodic function of $\l$
with period $i\pi$.
Notice also that $\Lambda_D (\pm \Th , {\vec \l}) = 0 $. That is, only
$\Lambda_A (\pm \Th , {\vec \l}) $ contributes to
the energy and momentum eigenvalues:
$$\eqalign{
         E(\Th ) &= a^{-1}\sum_{j=1}^M  \left[
      \phi(\Th+\l_j,\g/2) + \phi(\Th-\l_j,\g/2) - 2\pi \right] \cr
         P(\Th ) &= a^{-1}\sum_{j=1}^M  \left[
      \phi(\Th+\l_j,\g/2) - \phi(\Th-\l_j,\g/2) \right] \cr}    \eqn\enmom
$$
The ground state and the particle--like excitations of the light--cone
six--vertex model are well known [\ddv,\rev]: the ground state corresponds to
the unique solution of the BAE with $N/2$ consecutive real roots
(notice that the energy in eq.\enmom\ is negative definite, so that
the ground state is obtained by filling the interacting Dirac sea). In
the limit $N\to\infty$ this yields the antiferromagnetic vacuum. Holes
in the sea appear as physical particles. A hole located at $\varphi$
carries energy and momentum, relative to the vacuum,
$$
   e(\varphi)=2a^{-1}\arctan\left({{\cosh{\pi\varphi/\g}}
          \o{\sinh{\pi\Th/\g}}}\right)          \;,\quad
   p(\varphi)=-2a^{-1}\arctan\left({{\sinh{\pi\varphi/\g}}\o
            {\cosh{\pi\Th/\g}}}\right)                      \eqn\enmomi
$$
In the scaling limit $a\to0,\;\Th\to\infty$ with $e(0)$ held fixed, we
then obtain $(e,p)=m(\cosh\t,\sinh\t)$ with
$$
 m \equiv 4a^{-1}\exp(-\pi\Th/\g)\;,\quad  \t\equiv-\pi\varphi/\g  \eqn\massa
$$
identified,
respectively, as physical mass and physical rapidity of a sG soliton
(mT fermion) or antisoliton (antifermion). Complex roots of the BAE
are also possible. They correspond to {\it magnons}, that is to
different polarization states of several sG solitons
(mT fermions) , or to breather states
(in the attractive regime $\g>\pi/2$). In the rest of this paper, we shall
restrict our attention to the repulsive case $\g<\pi/2$, where the complex
roots corresponding to the breathers are absent.

\chapter{ Thermodynamic limit of the transfer matrix}

We proceed now to evaluate the function $G(\l,{\vec\l}\,)$ in the infinite
volume limit ($N\to\infty$ at fixed lattice spacing) for the
antiferroelectric ground state (the physical vacuum) and for
the excited states, in the repulsive regime $\g<\pi/2$.

For the vacuum, the density of roots $ \vec{\l}_V$  results to be [\jaca]
$$
\rho(\l)_V = N\int_{-\infty}^{+\infty} {{dk}\o{2\pi}} e^{ik\l} {{\cos{k\Th}}
\o {\cosh{k\g/2}}}
                                    \eqn\denv
$$
Using the integral representation
$$
       \phi(\l,\g/2) = {\rm P}\!\! \int_{-\infty}^{+\infty} {{dk}\o{ik}}
       e^{ik\l} {{\sinh{k\left[\pi/2-\g \right]}}\o {\sinh{k\pi/2}}}
\eqn\fii
$$
which is valid for $|\Im\l | < {\g\o2}$, and eq.\denv, we obtain
for $G(\l)_V \equiv G(\l,{\vec\l}_V)$
$$
G(\l)_V = -iN  \,{\rm P}\!\! \int_{-\infty}^{+\infty} {{dk}\o {k}} e^{ik\l}
{{\cos{k\Th} \sinh{k(\pi-\g)/2}}\o {\cosh{k\g/2} \sinh{k\pi/2}}}
\quad , \quad |\Im\l | < \g /2                              \eqn\gvi
$$
When  $\pi - \g/2 > |\Im\l | > {{\g}\o{2}}$  we need another
integral representation for $ \phi(\l,\g/2)$,
$$
     \phi(\l,\g/2) = -  {\rm P}\!\! \int_{-\infty}^{+\infty}
      {{dk}\o {ik}} e^{ik\l+(\pi k/2) \sign(\Im\l) }
              {{\sinh{k\g/2}}\o {\sinh{k\pi/2}}}   \eqn\fiotr
$$
We then find using eqs.\gee,\denv\ and \fiotr,
$$
G(\l)_V = iN \, {\rm P}\!\!\int_{-\infty}^{+\infty} {{dk}\o {k}}
      e^{ik\l+(\pi k/2) \sign(\Im\l)}
{{\cos{k\Th} \,\sinh{k \g/2}}\o {\cosh{k\g/2} \,\sinh{k\pi/2}}}      \eqn\gvf
$$
when  $\pi - \g/2> |\Im\l | > \g /2$.
That is, the function $G(\l)_V$ is discontinuous on the lines
$ \Im\l = \pm\g/2 $. As we shall see this fact holds true also for all
excited states. On the other hand  $G(\l,{\vec\l}\,)$ is periodic with
period $i\pi$, so that there exist two main analytic
determinations of its infinite volume limit $G(\l)$, that
we shall call henceforth  $G^I(\l)$ and $G^{II}(\l)$.
$$\eqalign{
     G(\l) &= G^I(\l) \; , \; {\rm for \, the \, strip \, I} :
       \quad -\g/2 < \Im\l < \g/2 \cr
        G(\l) &= G^{II}(\l) \; ,\; {\rm for \, the \, strip \, II} :
         \quad -\pi + \g/2 < \Im\l < -\g/2 \cr} \eqn\zona
$$
$G^I(\l)_V$ and $G^{II}(\l)_V$ have,
respectively, the integral representations \gvi\ and \gvf.
The functions $G^I(\l)_V$ and $G^{II}(\l)_V$ analytically continued in
$\l$ are meromorphic functions. Of course, they do not coincide with
$G(\l)$ except for the strips indicated in eq. \zona\ . For
$\Im\l$ outside these two strips, $G(\l)$ can be expressed in terms of
 $G^I(\l)_V$ or $G^{II}(\l)_V$ using the $i\pi$ periodicity as
follows:
$$\eqalign{
 G(\l) &= G^I(\l-in\pi) \quad {\rm for} \quad n\pi-\g/2 <\Im\l<n\pi+\g/2 \cr
 G(\l) &= G^{II}(\l-in\pi) \quad {\rm for}\quad
          (n-1)\pi+\g/2<\Im\l< n\pi-\g/2 \cr}         \eqn\banda
$$
where $n\in\ZZ$. The reflection principle also holds here:
$$
 G(\l) =  {\bar G({\bar \l})}
$$
We find from eqs.\gvi\ and \gvf\ the following expression for the difference
between the meromorphic functions $G^I(\l)_V$ and $G^{II}(\l)_V$:
$$
 G^{II}(\l)_V - G^{I}(\l)_V = -2iN {\rm Arg}\tanh\!\left[{{\cosh\left(
\pi\Th/\g\right)}\o{\cosh\left(\pi\l/\g\right)}}\right]
\eqn\disgv
$$
The discontinuities of $G(\l)$ through the other cuts follow by $i\pi$
periodicity and the reflection principle.

In addition, when $\l$ and  $\l-i\g$  lay both in
strip II (which is indeed possible for $\g<\pi/2$), we can relate
the functions $G(\l)_V $ and $G(\l - i \g)_V $ as follows :
$$\eqalign{
    G^{II}(\l)_V &+ G^{II}(\l -i \g)_V =
                      - iN \, {\rm P}\!\! \int_{-\infty}^{+\infty} {{dk}\o {k}}
     e^{ik\l-\pi k/2} ( 1 + e^{k\g} )
          {{\cos{k\Th}\,\sinh{k \g/2}}\o {\cosh{k\g/2}\,\sinh{k\pi/2}}} \cr
                 &= N \left[ \phi(\l-i\g/2 - \Th , \g /2) +
                    \phi(\l-i\g/2 + \Th, \g /2) \right]\cr}  \eqn\gvrel
$$
We find an analogous relation when $\l$ lays in strip I  and
$\l-i\g$  in strip II
$$\eqalign{
      G^I(\l)_V + G^{II}(\l - i \g)_V &=
                   N \left[ \phi(\l-i\g/2 - \Th , \g /2) +
                \phi(\l-i\g/2 + \Th, \g /2) \right] \cr
                               &+ iN \log{{\cosh{{\pi\Th}\o {\g}}
            - i \sinh{{\pi \l}\o{\g}}}\o{\cosh{{\pi\Th}\o {\g}}
            + i \sinh{{\pi \l}\o{\g}}}}  \cr}                    \eqn\gvred
$$
 Let us now consider excited states. We start with a two hole
state (the number of holes is always even when $N$ is even). The
density of roots is then [\jaca ]
$$
\rho (\l ) = \rho(\l)_V +  \rho(\l - \varphi _1)_h + \rho(\l -
\varphi_2)_h - \delta(\l - \varphi_1) - \delta(\l - \varphi_2)
\eqn\exden
$$
where $\varphi_1$ and $\varphi_2$ are the hole positions and
$$
\rho(\l)_h= \int_{-\infty}^{+\infty} {{dk} \o {2 \pi}}{{ e^{ik\l}
\sinh[{k\o 2}(\pi - 2 \g )]} \o { \sinh{{k \pi} \o 2} +
\sinh[{k\o 2}(\pi - 2 \g )]}}
\eqn\funp
$$
The function $G(\lambda)$ takes then the form
$$
G(\l) = G(\l)_V + G(\l - \varphi _1)_h + G(\l -\varphi_2)_h \eqn\gexci
$$
We find from eqs.\gee, \fii, \exden, and \funp\
$$\eqalign{
  G^I(\l)_h &= i \,{\rm P}\!\!\int_{-\infty}^{+\infty}{{dk} \o k}{{e^{ik\l}
     }\o{2\cosh{{k\g}\o 2}}} = -2\arctan\left(\tanh\frac{\pi\l}{2\g}\right)\cr
  G^{II}(\l)_h &= -i \,{\rm P}\!\!\int_{-\infty}^{+\infty}{{dk}\o{2k}}{{e^{ik\l
           + k (\pi /2)\sign(\Im\l)}
\sinh{{{k \g}\o 2}}}\o{\cosh{{k\g}\o 2} \sinh[{k \o 2}(\pi - \g)]}}\cr
 &= -2\arctan\left(\tanh\frac{\pi\l}{2\g}\right)
     +i\log S(\pi\l/\g - \sign(\Im\l)i\pi/2)  \cr} \eqn\fudos
$$
where $S(\t)$ is recognized as the soliton--soliton scattering
amplitude \amp\ upon the identification
$$
          {\hat \g} \equiv {\g \o {1 - {\g \o \pi}}}           \eqn\renor
$$
The function $S(\t)$ enjoys the crossing property
$$
       S(i\pi - \t) = {\hat b}(\t)~S(\t)          \eqn\cruce
$$
where
$$
{\hat b}(\t) = {{\sinh({{{\hat \g} \t}\o\pi})}\o{\sinh{{\hat \g}\o\pi}
(i\pi - \t)}}                                        \eqn\be
$$
Notice that ${\hat b}(\t)~{\hat b}(i\pi - \t) = 1$ .
We see from eq.\fudos\  that $G(\l)_h$ has cuts on
the lines $\Im\l = \pm \g/2$, with discontinuity
$\quad i\log S({\pi\o\g}\l \mp i{\pi\o 2})$.

Next consider states containing complex roots. There are four
kinds of complex roots [\woyn] associated to excited states close
to the $N\to\infty$
antiferromagnetic vacuum, in the regime $\g<\pi/2$:
\item{a)}
Close roots with $|\Im \l | < \g $ . They appear as quartets :
$\l = (\s \pm i \eta ,\, \s \pm i[\g-\eta]), $ where $ 0 < \eta < \g $ ,
or as two strings: $\l = \s \pm i\g/2 $.
\item{b)}
Wide roots with $|\Im\l|>\g$. They appear in pairs
$\l = \s \pm i\eta$, $\g/2<\eta<\pi/2$,
or as self--conjugate single roots with $|\Im\l|=\pi/2$.

The presence of such complex roots produces a backflow in the real roots
density. For a close pair we have [\qgba ]
$$
\rho_\eta(\l)_c = -{1 \o {2\pi}}[ p(\l-\s-i\eta)+ p(\l-\s+i\eta)] \;,
      \quad                   \eta<\g<\pi/2                    \eqn\dcp
$$
while for a wide pair [\qgba ]
$$
\rho_\eta(\l)_w= -{1 \o {2\pi}}{d\o{d\l}}[\phi_{\g}(\l-\s ,\eta - \g) -
\phi_{\g}(\l-\s ,\eta )]     \;,\quad  \eta>\g<\pi/2              \eqn\dwp
$$
where
$$
            \phi_{\g}(\l,\eta)\equiv
          \phi\bigl({\l\o{1-\g/\pi}},{\eta\o{1-\g/\pi}}\bigr)
\eqn\fig
$$
A self--conjugate root at $\s+i\pi/2$ gives instead
$$
     \rho(\l)_{sc}= \frac12 \rho_{\pi/2}(\l)_w               \eqn\wtosc
$$
Let us denote by $G_\eta(\l)_c$ and $G_\eta(\l)_w$ the contribution of
a closed pair and of a wide pair to the function $G(\l)$,
respectively. For self--conjugate roots one simply has
$G(\l)_{sc}=\frac12 G_{\pi/2}(\l)_w$. We find from eqs. \gee\ and \dcp\ :
$$
 G_\eta(\l)_c = \int_{-\infty}^{+\infty}d\mu \, \phi(\l - \mu , \g/2)
  \rho_\eta(\mu)_c
+ \phi(\l - \s - i\eta, \g /2)+ \phi(\l - \s + i\eta, \g /2)
\eqn\gcp
$$
Then the integral representations \fii\ for $\phi(\l,\g /2)$
and the density \dcp\ yield
$$
    G^I_\eta(\l)_c =2\arctan\left(\tanh\frac{\pi}{2\g}[\l-\s-i\eta]\right)
     +   2\arctan\left(\tanh\frac{\pi}{2\g}[\l-\s+i\eta]\right)  \eqn\men
$$
It is easy to see that the total contribution for a quartet vanishes
when $ |\Im \l| < \g/2 $ and that the two--string contributions equal
$\pm \pi$ in this region:
$$
  G^I_\eta(\l)_c +  G^I_{\g-\eta}(\l)_c = 0 \;{\rm mod}\,2\pi, \quad
      G^I_{\g/2}(\l)_c= i\log(-1)  \;,\quad      |\Im \l | < \g/2   \eqn\dues
$$
Hence, quartets and two--strings do not contribute to the energy and
momentum.

Let us now consider the more interesting strips of type II.
There, using the integral representation \fiotr\ for
$\phi(\l, \g /2)$, we obtain
$$\eqalign{
 G^{II}_\eta(\l)_c &= 2\arctan\left(\tanh\frac{\pi}{2\g}[\l-\s-i\eta]\right)
         + 2\arctan\left(\tanh\frac{\pi}{2\g}[\l-\s+i\eta]\right) \cr
  & -i\log S\bigl(\frac\pi\g [\l-\s +i \eta] -i\frac\pi2 \bigr)
    -i\log S\bigl(\frac\pi\g [\l-\s -i \eta] -i\frac\pi2 \bigr) \cr}\eqn\mas
$$
Then, the total contribution for quartets and two strings results in
$$\eqalign{
 G^{II}_\eta(\l)_c +  G^{II}_{\g-\eta}(\l)_c = & \;i \log \bigl[
{\hat b}\bigl(\frac\pi\g [\l-\s +i \eta] -i\frac\pi2 \bigr)
{\hat b}\bigl(\frac\pi\g [\l-\s -i \eta] -i\frac\pi2 \bigr)\bigr] \cr
 G^{II}_{\g/2}(\l)_c= & \;i \log\bigl[-
  {\hat b}\bigl(\frac\pi\g [\l-\s ]\bigr)\bigr] \cr}   \eqn\totc
$$
where we used eq.\cruce.

Let us finally consider the wide pairs. Their contribution is given
by
$$
 G_\eta(\l)_w = \int_{-\infty}^{+\infty} d\mu \, \phi(\l - \mu , \g/2)
    \rho_\eta(\mu)_w
+ \phi(\l - \s - i\eta, \g /2)+ \phi(\l - \s + i\eta, \g /2)   \eqn\gwp
$$
Use of eqs. \fii\ , \fiotr\ and \dwp\ now yields
$$\eqalign{
             G^I_\eta(\l)_w &= 0  \quad {\rm mod}\,2\pi    \cr
             G^{II}_\eta(\l)_w & = \phi_{\g}(\l-\s - i\g/2 ,\g - \eta) +
          \phi_{\g}(\l-\s - i\g/2 ,\eta ) \cr}              \eqn\anc
$$
We see from eq.\anc\ that wide pairs do not contribute to the energy and
momentum.

We are now in position to analyze the excitation spectrum of the
infinite--volume transfer matrix $t(\l,\Th)$. Let us begin with $\l$
lying in strip I. From eqs. \llamm\ and \gvred, we find for the vacuum
$$ \eqalign{
\Lambda_A(\l)_V &= \exp\left[-iG^I(\l)_V\right]    \cr
\Lambda_D(\l)_V &= \exp\left[-iG^I(\l)_V\right]
           \left({{\cosh{\pi\Th/\g}
         + i \sinh{\pi\l/\g}}\o{\cosh{\pi\Th/\g}
         - i \sinh{\pi\l/\g}}}\right)^N         \cr}              \eqn\autv
$$
The extra factor in $\Lambda_D(\l)$ tends to zero (infinity)
for $\Im \l$ positive (negative) when $N \to \infty$.
Since this vacuum contribution is present in any physical
particle--like excitation, we see that the $\Lambda_D(\l)$ will always
behave like $\Lambda_D(\l)_V$. Let us recall that $\Lambda_D(\pm\Th)=0$
for any finite $N$, giving no contribution to energy and
momentum. Hence, choosing $0<\Im\l<\g/2$, we are guaranteed that the two
limits $\l\to\pm\Th$ and $N \to \infty$ commute. The reduced strip
$0<\Im\l<\g/2$ is therefore the most natural one to define the
renormalized type I transfer matrix
$$
    t^I(\l)=  \lim_{N\to\infty} t(\l,\Th)
             \exp\bigl[iG^I(\l)_V\bigr](-)^{J_z-N/2}  \eqn\rentm
$$
where $J_z=N/2-M$ is to be identified with the soliton (or fermion) charge
of the continuum sG--mT model. The last sign factor in eq.\rentm\ corresponds
to square--root branch choice suitable to obtain the relation
$$
      t^I(\pm\Th)= \exp\{-ia[P_{\pm}-(P_{\pm})_V]\}         \eqn\tppm
$$
where $ P_{\pm} \equiv (H\pm P)/2$ (see eqs.\evolu, \ttou, \enmom) and
$(P_{\pm})_V$ stands for the vacuum contribution.
Notice that the $\Th-$dependence of  $t^I(\l)$ has been completely
canceled out, since it is present only in the vacuum contribution. In fact,
from eqs.\gexci, \fudos, \dues\ and \anc, we read the eigenvalue
$\Lambda^I(\l)$ of $t^I(\l)$ on a generic particle state:
$$
     \Lambda^I(\l)= \exp\left[-2i\sum_{n=1}^k \arctan
              \left(e^{\pi\l/\g+\t_n} \right)\right] =  \prod_{n=1}^k
       \coth\left({{\pi\l}\o{2\g}}+{{\t_n}\o2}+{{i\pi}\o 4}\right)   \eqn\eigth
$$
where $\t_n\equiv -\pi\varphi_n/\g$ are the physical particle rapidities.
Suppose now we expand  $\log{\Lambda^I}(\l)$ in
powers of $z=e^{-\pi|\l|/\g}$ around $\l=\pm\infty$,
$$
  \pm i \log\Lambda^I(\l) =   \sum_{j=0}^\infty z^{2j+1}
                {{(-1)^j}\o{j+1/2}} \sum_{n=1}^k e^{\pm(2j+1)\t_n}  \eqn\expa
$$
One has to regard the coefficents of the expansion parameter $z$ as the
eigenvalues of the conserved abelian charges generated by the transfer matrix.
The additivity of the eigenvalues implies the locality of the charges.
In terms of operators we can write, around $\l=\pm\infty$,
$$
    \pm i \log t^I(\l)  = \sum_{j=0}^\infty
                 \left[{{4z}\o m}\right]^{2j+1} I_j^\pm     \eqn\expaop
$$
where $I_0^\pm = p_{\pm} $ is the continuum light--cone energy--momentum and
the  $I_j^\pm$, $j \ge 1$, are local conserved charges  with dimension $2j+1$
and  Lorentz spin $\pm(2j+1)$. Their eigenvalues
$$
       {{(-1)^j}\o{j+1/2}} \sum_{n=1}^k\left[{m\o4}e^{\pm\t_n}\right]^{2j+1}
$$
coincide with the values on multisoliton solutions of the
higher integrals of motion of the sG equation [\leon].
It is remarkable that these eigenvalues are free
of quantum corrections although the corresponding operators in terms of
local fields certainly need renormalization. Let us stress that
explicit expressions for these conserved charges can be obtained
by writing the local $R-$matrices in terms of fermi operators, as in
ref.[\ddv]. Notice also that, combining
eqs.\tppm\ with \expaop, and recalling the scaling law \massa, we can write
$$
     P_{\pm}-(P_{\pm})_V = p_{\pm}+{m\o4}\sum_{j=1}^{\infty}
               \left({{ma}\o4}\right)^{2j} I_j^\pm       \eqn\hmpl
$$
That is, the light-cone lattice hamiltonian and momentum can be expressed in a
precise way as the continuum hamiltonian and momentum plus an infinite series
of continuum higher conserved charges, playing the r\^ole of irrelevant
operators.

We now come back to the problem of comparing the light--cone results with the
bootstrap predictions. As we have just seen, there is no chance to match the
bootstrap predictions  for $\l$ in strip I, since $\Lambda_A(\l)$ and
$\Lambda_D(\l)$ cannot be renormalized by a common factor (see eq.\autv).
Indeed,  the structure of the sum $\Lambda_A(\l)+\Lambda_D(\l)$, that is of the
eigenvalue $\Lambda(\l)$ of $t(\l,\Th)$, will never match that of the
eigenvalue $\xi(u)$ of the bootstrap transfer matrix $\tau(u)$ (eq.\heig). The
situation is more favourable when both $\l$ and $\l-i\g$ lay in
strip II. In this case eq.\gvrel\ applies and we find that
$$
\Lambda_A(\l)_V= \Lambda_D(\l)_V = \exp\left[-iG^{II}(\l)_V\right]  \eqn\autvo
$$
In order to consider all other excited states, it is important to recall that
in the infinite volume limit the complex roots and the holes are coupled by
equations with the BA structure [\woyn]. These ``higher--level'' BAE follow
from the original BAE, eq.\bae, by summing up the Dirac sea of real roots in
much the same way as we have done here for the function $G(\l)$. The result can
be cast in the most symmetrical form by parametrizing the complex roots as
follows [\woyn]:
\item{a)}  $\s={\g\o\pi} u$, for two--strings
\item{b)}  $\s+i\eta ={\g\o\pi}(u+i\pi/2)$ and
$\s-i\eta={\g\o\pi}({\bar u}-i\pi/2)$,  for quartets and wide pairs.
\item{c)}  $\s+i\pi/2 ={\g\o\pi}(u+i\pi/2)$ for self--cojugate roots.

Then the equations satisfied by the new complex root parameters
$\{u_j,\;j=1,2,\ldots,m\}$ exactly coincide with the bootstrap BAE
\bae, upon the natural identification of $-\pi\varphi_n/\g$ with the
physical rapidity $\t_n$ of the $n$th hole (or particle) where $1 \leq
n \leq k$ .
By construction, the number $m$ of higher--level roots is equal to
the number of two--strings and self--conjugated roots plus twice the
number of quartets and wide pairs. Notice that a self--conjugate
root in the bare BAE is also self--conjugate in the higher--level
BAE.

Then, combining eqs.\totc, \anc, \autvo\ and using the new
$u-$parametrization for the complex roots,
we obtain the general form of the $A$ and $D$ contributions to the
eigenvalue of $t(\l,\Th)$ on the $N\to\infty$ limit of the BA states for
$\l$ in strip II:
$$
     \Lambda_A(\l) = -e^{-iG^{II}(\l)_V} \left\{ \prod_{n=1}^k
         S(x_n) \coth{{x_n}\o2}  \right\}
        \prod_{j=1}^m
             {{\sinh {\hat\g} [i/2+({\pi\o\g}(\l+i\g/2)-u_j)/\pi]}  \o
              {\sinh {\hat\g} [i/2-({\pi\o\g}(\l+i\g/2)+u_j)/\pi]}}  \eqn\lsuba
$$
and $$
 \Lambda_D(\l) = -e^{-iG^{II}(\l)_V} \left\{ \prod_{n=1}^k
        S(x_n){\hat b}(x_n) \coth{{x_n}\o2}\right\}
        \prod_{j=1}^m
             {{\sinh {\hat\g} [3i/2+({\pi\o\g}(\l+i\g/2)-u_j)/\pi]} \o
              {\sinh {\hat\g} [-i/2+({\pi\o\g}(\l+i\g/2)-u_j)/\pi]}} \eqn\lsubd
$$
where for definiteness we chose the strip II , $-\pi+\g/2<\Im\l<-\g/2$ and set
$x_n={\pi\o\g}(\l+i\g/2)+\t_n$.
These last two expressions can be connected  with that for the eigenvalues of
the bootstrap transfer matrix $\tau(u)$, eq.\heig, provided we {\it identify}
$u$ with ${\pi\o\g}(\l+i\g/2)$. We find indeed from eqs.\heig, \eigev,
\lsuba, \lsubd :
$$
          \Lambda(\l) = -e^{-iG^{II}(\l)_V} \xi(u) \prod_{n=1}^k
          \coth\left({{u+\t_n}\o2}\right)        \eqn\landxi
$$
with $\l$ in strip II . In analogy with eq.\rentm, we now define the type
II renormalized transfer matrix
$$
  t^{II}(\l) =  \lim_{N\to\infty} t(\l,\Th)
        \exp\bigl[iG^{II}(\l)_V\bigr](-)^{J_z-N/2}  \eqn\rentmb
$$
Then, taking into account eq.\eigth, eq.\landxi\ can be rewritten
$$
  \xi(u) =  {{\Lambda^{II}\left({\g\o\pi}u - i{\g\o2}\right)}
              \o {\Lambda^I\left({\g\o\pi}u - i{\g\o2}\right)}}     \eqn\xilan
$$
Notice that the dependence on the cutoff rapidity $\Th$
has completely disappeared from the r.h.s. of eq.\xilan. This holds true both
for the explicit dependence in the vacuum function $G(\l)_V$ and
for the implicit dependence through the bare BAE, which are now
replaced by the $\Th-$independent higher--level ones. In other words, the
eigenvalues of the bootstrap transfer matrix can be recovered from the
light--cone regularization already on the infinite diagonal lattice, with no
need to take the continuum limit.
This should cause no surprise, since after all a factorized
scattering can be defined also on the infinite lattice, with physical
rapidities replaced by lattice rapidities (see eq.\enmomi). The bootstrap
construction of the quantum monodromy operators $\T_{ab}(u)$ then proceeds just
like on the continnum. In this case, some $q_0-$deformation of the two
dimensional Lorentz algebra should act as a symmetry on the physical
states. This $q_0$ becomes unit when $\Th \to \infty$.

\chapter{ Final remarks }

In the previous section we have established the precise relation \xilan\
between the BA eigenvalues of the bootstrap and microscopic lattice
transfer matrices
sG--mT--6V model, when  $|\Im u|<\pi/2$ and  $\g<\pi/2$.
With the implicit understanding that the thermodynamic limit $N\to\infty$
is taken in the ground state representation, such a relation extends
to the operators themselves:
$$
     \tau(u)= t^{II}\bigl({\g\o\pi}u-i{\g\o2} \bigl)
                   \,t^I\bigl({\g\o\pi}u-i{\g\o2}\bigl)^{-1}   \eqn\tauet
$$
where $ \tau(u)$ is the bootstrap operator \trans\ . The relation
\tauet\ between $\tau(u)$ and  $t(\l,\Th)$ is remarkably simple,
specially taking into account the long chain of steps involved in their
totally independent constructions. For $t(\l,\Th)$ we have:
\item{1.} Defined the light-cone lattice with the alternating parameter $\Th$.
\item{2.} Found the antiferroelectric ground state.
\item{3.} Considered general finite energy excitations around it.
\item{4.} Let the volume $N$ become infinity.

In the other hand,  $\tau(u)$ follows solely from the bootstrap principles
(a)--(c) of sec. 2.

Notice that
the bootstrap construction by itself does not provide any relationship
between $\T_{ab}(u)$ and the local fundamental fields entering the lagrangian
which supposedly corresponds to the given factorized scattering  model.
On the other hand $T_{ab}(\l,\Th)$ can be explicitly written in terms of the
bare fermi field of the mT--model [\ddv], so that eq.\tauet\ represents a
relevant piece of information for the search of such a relationship.
It is clear, however, that a direct extension of eq.\tauet\ to the full
monodromy matrix would not work: indeed, suppose that operators
${\tilde\T}_{ab}(u)$ are consistently defined by the relation
$$
   {\tilde\T}_{ab}(u)=
                   T_{ab}^{II}\bigl({\g\o\pi}u-i{\g\o2},\Th\bigl)
                   \,t^I\bigl({\g\o\pi}u-i{\g\o2}\bigl)^{-1}   \eqn\TaueT
$$
then certainly the trace $\sum_a {\tilde\T}_{aa}(u)$ coincides with
$\tau(u)$, due to eq.\tauet, but ${\tilde\T}_{ab}(u)$ cannot
be identified with $\T_{ab}(u)$ because it still satisfies a bare YB algebra,
with anisotropy $\g$ rather than $\hat\g$. [In the YB algebra \yba\ the
R-matrix elements, as given by eq.\amp, depend on $\hat\g$ ].
 It is presumable therefore
that eq.\TaueT\ does not provide a consistent renormalization for the
complete monodromy matrix. It should be noted, in this respect, that
all the models considered in refs.[\dema], where the existence of a
classsical analogue of $\T_{ab}(u)$ allows to relate it to
local curvature--free divergenceless non--abelian currents, correspond to
rational forms of the $R-$matrix. But then one would find no finite
renormalization like $\g\to {\hat\g}$ when taking the $N\to\infty$ limit in the
light--cone lattice regularization of these models. Since both bare and
bootstrap $R-$matrices are rational and depend non--trivially only on the
spectral parameter,  it is always possible to rescale the latter so that
bare and
boootstrap YB algebras coincide. In other words, in these rational models,
there
exist a thermodynamic limit in which the microscopically defined lattice
monodromy matrix is  conserved. Notice that this lattice monodromy matrix can
be written in term of lattice non--abelian currents [\ddva ] in a way which
represents an {\it integrable} regularization of the classical monodromy
matrix. The picture is therefore fully consistent for the rational models.

Evidently, the situation appears to be more subtle in a trigonometric
integrable model like the sG--mT--6V model considered here in detail.
At the microscopic level the model enjoys a dynamical YB symmetry
characterized by the anisotropy $\g$, which underlies the BA solution based on
the ferromagnetic reference state $\Omega$.
At the ``renormalized" level, when the reference state is the physical
antiferromagnetic ground state of the infinite lattice (and still in the
presence of the UV cutoff provided by the lattice spacing), the model
aquires a true YB symmetry characterized by the anisotropy $\hat\g$.
Eq.\tauet\ shows that the Cartan subalgebras of these two YB algebras
are essentially identical, strongly supporting both the bootstrap and
the light--cone lattice constructions. It would be very interesting to
relate the complete monodromy matrices, that is to find general YB--algebraic
arguments to provide a microscopic interpretation for the bootstrap monodromy.
The recent work reported in refs. [\japs], which relies on the $q-$deformed
affine algebra approach to the YB symmetry, seems very promising in this
respect, although it is restricted to the regime $|q|<1$
(while $|q|=1$ in the sG--mT--6V model).

\refout

\bigskip
\bigskip

\centerline{FIGURE CAPTIONS}
\medskip
\item{Fig.1.}
Light--cone lattice representing a discretized portion of
Minkowski space--time. An $R-$matrix of probability amplitudes is attached
to each vertex. The bold lines correspond to the action, at a given time,
of the one--step evolution operator $U$.
\item{Fig.2.}
Graphical representation of the inhomogeneous monodromy matrix. The angles
between the horizontal and the vertical lines are site--dependent in an
arbitrary way.
\item{Fig.3.}
Insertion of the alternating monodromy matrix in the light--cone lattice.
\item{Fig.4.}
The two main determinations, $G^I(\l)$ and $G^{II}(\l)$ are defined by
$G(\l)$ with $\l$ in strips I and II, respectively.

\bye
The third possibility, $\l$ in a type II strip and $\l-i\g$ in
a  type I strip, can be related to one of the previous two by
periodicity. We consider the
functions $G^I(\l)$ and $G^{II}(\l)$ extended by analiticity
to  the whole complex plane. They are meromorphic functions
of $\l$.
\bye